\documentclass{optica-article}

\journal{opticajournal} 

\articletype{Research Article}

\usepackage{lineno}
\usepackage{graphicx}
\usepackage{amsmath}
\usepackage{mathtools}
\usepackage{braket}
\usepackage[title]{appendix}

\begin{document}

\title{Hong-Ou-Mandel test to verify indistinguishability of the states emitted from a quantum key distribution transmitter implementing decoy Bennett-Brassard 1984 protocol}

\author{Shunya Tajima,\authormark{1,3} Akihisa Tomita,\authormark{2,4,*} and  Atsushi Okamoto\authormark{2}}

\address{\authormark{1} Graduate School of Information Science and Technology, Hokkaido University, Sapporo 060-0814, Japan\\
\authormark{2} Faculty of Information Science and Technology, Hokkaido University,  Sapporo 060-0814, Japan\\
\authormark{3} Currently with NEC Corporation, Japan\\
\authormark{4} Currently with National Institute of Information and Communications Technology, Tokyo, Japan}

\email{\authormark{*}a.tomita@nict.go.jp} 


\begin{abstract*} 
Quantum Key Distribution (QKD) systems require rigorous verification of device properties to ensure implementation security. A critical requirement is the indistinguishability of transmitted pulses encoded by different modulation patterns, as distinguishability through non-encoded degrees of freedom could enable undetected eavesdropping. We present a practical method for testing pulse indistinguishability in QKD transmitters based on Hong-Ou-Mandel (HOM) interference. We establish the theoretical equivalence between the SWAP test and HOM measurement for characterizing quantum state fidelity, demonstrating that HOM visibility directly relates to the trace of density matrix products for phase-randomized weak coherent pulses. We experimentally validated this approach using a high-speed QKD transmitter implementing the decoy BB84 protocol with time-bin encoding at 1.25 GHz. HOM interference was measured between adjacent pulses prepared in different Bennett-Brassard 1984 states (X0, X1, Y0, Y1) using superconducting nanowire single-photon detectors. The observed HOM visibility was approximately 0.3 across all state combinations, with no statistically significant differences detected. These results confirm that modulation does not compromise pulse indistinguishability in our transmitter. The HOM test provides a practical, quantum-optical method for security certification of QKD systems without requiring assumptions about specific degrees of freedom, using only standard fiber-optic components and single-photon detectors.

\end{abstract*}

\section{Introduction}
Quantum Key Distribution (QKD) provides an information-theoretically secure cryptographic key between two remote parties\cite{gisin2002quantum}. Since the introduction of the first QKD protocol, Bennett-Brassard 1984 (BB84)\cite{BB84}, the security of this protocol has been validated through proofs for both single-photon sources and weak coherent light sources\cite{lo_unconditional_1999,shor_simple_2000,mayers_unconditional_2001,hutchison_universal_2005,renner_information-theoretic_2005,koashi_simple_2009,Hwang2003_decoy,Wang2005_PRL_decoy,Lo2005_decoy}. 
However, these security proofs depend on several assumptions regarding the properties of the devices used\cite{GLLP,scarani2009security}. 
Therefore, it is essential to examine the properties of devices employed in practical QKD systems to ensure their security. 
This area is known as implementation security, which has recently garnered increased interest to promote the social deployment of QKD systems\cite{tomita_implementation_2019,xu_secure_2020,makarov_preparing_2024,zapatero_implementation_2025}. 
Researchers and providers of QKD systems are collaborating to establish standards on the evaluation method for practical QKD systems\cite{ISOIEC23837-2,etsi_industry_specification_quantum_2024}. 
Currently, these efforts are primarily focused on systems that implement the decoy BB84 protocol, which is widely adopted in most QKD systems.

For a transmitter, one of the most important properties to examine is the indistinguishability of encoded states. In a QKD transmitter, a photon is encoded into a qubit state required for the protocol by modulating a degree of freedom of the photon. 
If photons encoded to different states in the same degree of freedom can be distinguished by observing non-encoded degrees of freedom, an eavesdropper may obtain information on the encoded state without being detected. 
Recently, this type of flaw was analyzed by Gnanapansithan, \textit{et al.}\cite{gnanapandithan_hidden_2025}.
For example, most Quantum Key Distribution (QKD) systems that use optical fiber transmission employ time-bin quantum bits, which encode quantum information based on the relative arrival times of light pulses. This method is favored due to its resilience against depolarization and polarization mode dispersion occurring during transmission over fiber links.
Other degrees of freedom include transverse modes, wavelength, polarization, and arrival time. 
The evaluator should examine whether the pulses can be distinguished by measuring characteristics such as the spectrum and the temporal waveform. We may omit the transverse modes and polarization measurement because they can be defined in the transmitter simply by using single-mode polarization-maintaining fibers.

Methods for measuring the spectrum and waveform of classical light have been well established, as seen in a standard\cite{ISOIEC23837-2}. 
We can utilize commercially available tools such as an optical spectrum analyzer and an optical sampling oscilloscope. The indistinguishability can be investigated by estimating the fidelity of the probability distribution functions on non-encoded degrees of freedom for different encoded states.
However, measuring weak light at the single-photon level requires more sophisticated equipment, such as streak cameras. This increases the time and cost of measurement.
Moreover, it is not obvious that the spectrum and waveform measurements exhaust the non-encoded degrees of freedom. 
It is essential to develop a quantum optical method for examining the indistinguishability of optical pulses used in QKD communication.
The second- and fourth-order interference visibility was proposed as a means of investigating the side channels in practical QKD systems\cite{sych_practical_2021}.
However, a comprehensive investigation has not been conducted.

In this article, we analyze a method based on Hong-Ou-Mandel (HOM) interference\cite{hong_measurement_1987} and explore its experimental feasibility.
Our research investigated both theoretical and experimental approaches to characterizing indistinguishability in modulated photonic states.
In Section \ref{sec:theory}, we demonstrate the connection between the SWAP test and the HOM test, illustrating how HOM visibility can be used to characterize the fidelity of two quantum states. 
In Section \ref{sec:experiment}, we present the construction of the HOM test setup designed for high-speed quantum key distribution (QKD) transmitters. 
The QKD state generator utilized in our experiment was developed in our laboratory, and further details can be found in the appendix. 
In Section \ref{sec:results}, we present the results of the HOM test and conclude that encoding scarcely affected the HOM visibility in our experiment. 
In section \ref{sec:discussion}, we examine the experimental results. 
Finally, in Section \ref{sec:conclusion}, we conclude the article by making remarks on the proposed method.

\section{Theory}\label{sec:theory}
\subsection{Equivalence between SWAP test and Hong-Ou-Mandel test}
\begin{figure}[tbp] 
\centering
\includegraphics[width=\linewidth]{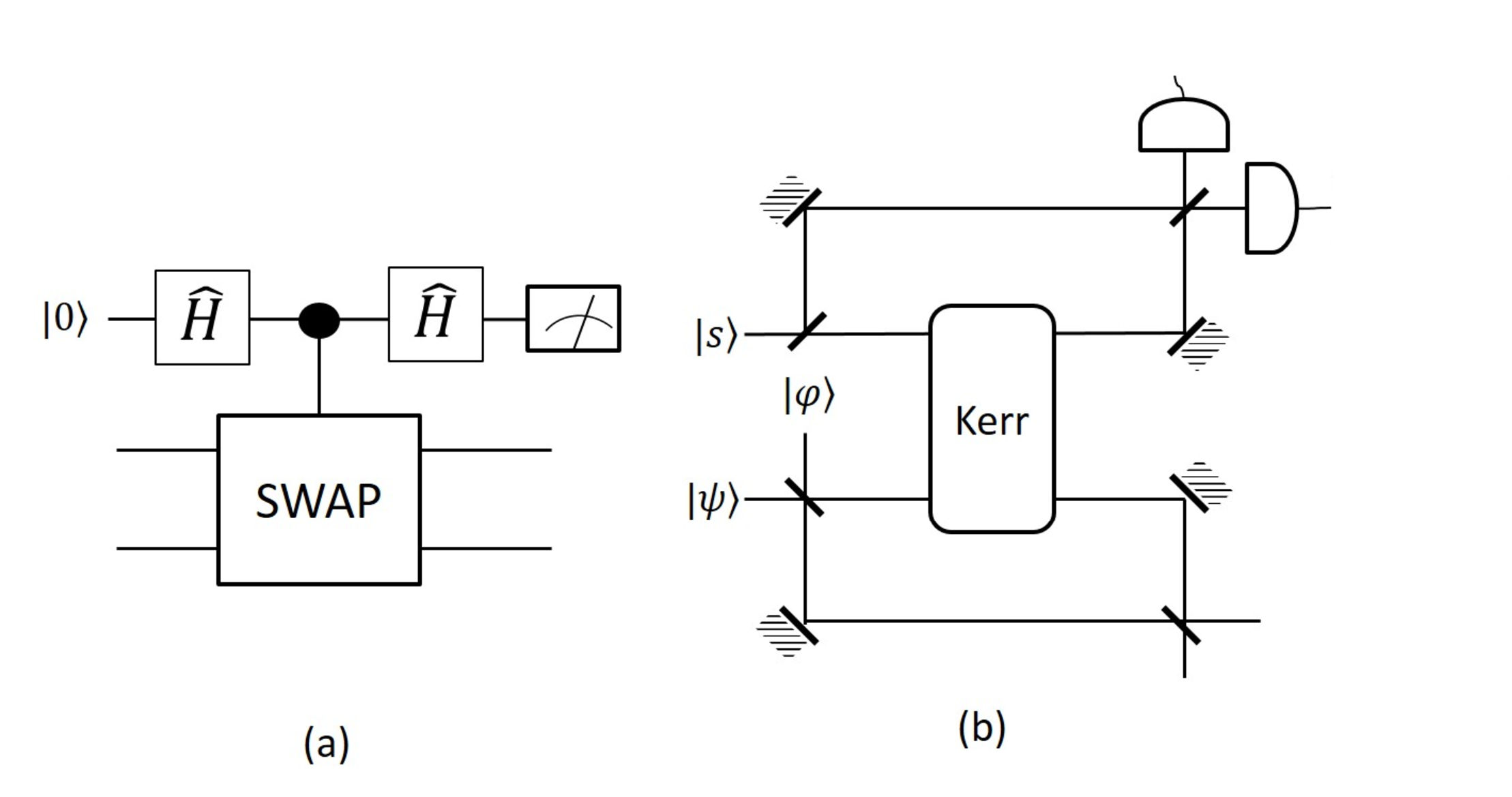}
\caption{Schematic of SWAP test. (a) Quantum circuit represented by quantum gates. (b) Photonic implementation. A controlled SWAP operation is constructed by a Mach-Zehnder interferometer (MZI) and Kerr nonlinearity. A single photon state $\ket{s}$ is the input for the upper MZI.}\label{fig:CSWAP}
\end{figure}

Fidelity between two pure states $\ket{\varphi}$ and $\ket{\psi}$ can be measured using the SWAP test, which was originally proposed as a sub-protocol of quantum fingerprinting\cite{buhrman_quantum_2001}.
Figure \ref{fig:CSWAP} (a) illustrates the quantum circuit for the SWAP test.  
A SWAP operation $\hat{U}_{\mathrm{S}}:\ket{\phi} \ket{\psi} \mapsto \ket{\psi} \ket{\phi}$ is performed conditionally on the control qubit,  which is prepared to be $\ket{+} = (\ket{0}+\ket{1})/\sqrt{2}$ by applying a Hadamard gate $\hat{H}$ to the initial state $\ket{0}$. 
After the controlled (c-)SWAP operation, a Hadamard gate is applied to the control qubit, which is then measured in the computational basis. 
The state after the second Hadamard gate operation is given by 
\begin{align}\label{eq:cSWAP}
    &(\hat{H} \otimes \hat{1}) \left(\ket{0}\bra{0} \otimes \hat{1} + \ket{1}\bra{1} \otimes \hat{U}_{\mathrm{S}}\right)
    (\hat{H} \otimes \hat{1})\ket{0}\ket{\phi}\ket{\psi} \nonumber\\
    =&\frac{1}{2} \left[\ket{0}(\ket{\phi}\ket{\psi}+\ket{\psi}\ket{\phi})+\ket{1}(\ket{\phi}\ket{\psi}-\ket{\psi}\ket{\phi})\right].
\end{align}
The probabilities of obtaining the measurement outcomes "0" and "1" are 
\begin{align}\label{eq:P_cSWAP}
     P(0) &= \frac{1+|\braket{\phi|\psi}|^2}{2} \nonumber \\
     P(1) &= \frac{1-|\braket{\phi|\psi}|^2}{2}.
\end{align}
Thus, the fidelity can be estimated as follows:
\begin{equation}\label{eq:fidelity}
    F(\ket{\phi,\ket{\psi}} = |\braket{\phi|\psi}| =\sqrt{\frac{P(0)-P(1)}{P(0)+P(1)} }.
\end{equation}
When the inputs are mixed states denoted by density operators $\rho_1$ and $\rho_2$, $|\braket{\phi | \psi}|^2$ in Eq. (\ref{eq:P_cSWAP}) should be replaced by $ \mathrm{Tr}(\rho_1 \rho_2)$\cite{buhrman_quantum_2001}.
Then RHS of Eq. (\ref{eq:fidelity}) no longer represents fidelity.
Rather, it relates to the trace of the product of the density operators as 
\begin{equation}\label{eq:trace_product}
    \mathrm{Tr}(\rho_1 \rho_2) = \frac{P(0)-P(1)}{P(0)+P(1)},
\end{equation}
which yields a lower bound and an upper bound of the fidelity as\cite{miszczak_sub-_2009}
\begin{equation}
    \mathrm{Tr}(\rho_1 \rho_2) \le F^2(\rho_1, \rho_2) \le \mathrm{Tr}(\rho_1 \rho_2) + \sqrt{S_{\mathrm{L}}(\rho_1) S_{\mathrm{L}}(\rho_2)},
\end{equation}
where $S_{\mathrm{L}}(\rho)$ is linear entropy defined by $S_{\mathrm{L}}(\rho)=1-\mathrm{Tr}(\rho^2)$.
The equality holds when the states are pure.
Even for the mixed states, the modulation effect on the indistinguishability can be expressed by the trace of the density-operator product given by the SWAP test. 
Unless the effects are identical for all the modulation patterns, the trace of the modulated density-operator product is less than that of the unmodulated density-operator product. 
The analysis using density operators is summarized in the appendix.

The SWAP test is suitable for our purpose because there is no restriction on the Hilbert space where the two states belong, as shown in the above derivation of the probabilities in Eq. (\ref{eq:P_cSWAP}).
The SWAP test can be implemented with a photonic circuit, as shown in Fig. \ref{fig:CSWAP} (b). 
Since a 50:50 beam splitter (HBS) acts as a Hadamard gate for a single photon state, a series of two Hadamard gates is realized by the upper Mach-Zehnder interferometer (MZI) in the figure.
The lower MZI provides a SWAP operation controlled by the phase difference between paths through Kerr nonlinearity. 
If the phase difference between the paths in the MZI is $\pi$, the two inputs are swapped, and the output appears unchanged if the phase difference is $0$.
However, the photonic implementation of the SWAP test is challenging because the c-SWAP operation requires a $\pi$ phase shift by a single photon, which implies significant optical nonlinearity. 
This problem can be circumvented by utilizing the equivalence of the SWAP test and HOM interference, as demonstrated by Garcia-Escartin and Chamorro-Posada\cite{garcia-escartin_swap_2013}.
The reduction of the c-SWAP to the HOM stems from the observation that the Kerr operation conserves the number of photons, as described by the operator 
\begin{equation}\label{eq:Kerr}
    \hat{U}_\mathrm{Kerr}=e^{i \pi \hat{n}_\mathrm{U} \hat{n}_\mathrm{L} }, 
\end{equation}
where the operators $\hat{n}_\mathrm{U}$ and $\hat{n}_\mathrm{L}$ stand for the number operators of the photons on the paths in the upper and the lower MZIs, respectively.
Therefore, if we allow the destruction of the input states $\ket{\varphi}$ and $\ket{\psi}$, we can replace the Kerr operation with a photon number measurement followed by a phase shift controlled classically on the upper MZI.
Once the photon number measurement is done, the results of the upper interferometer can be determined without measurement. If the photon number is odd, a $\pi$ phase shift should be applied to the upper MZI. This results in the outcome "1". Conversely, if the photon number is zero or even, no phase shift should be applied, resulting in the outcome "0". This observation allows for the exclusion of the upper MZI.
The construction of a test that is equivalent to the SWAP test is possible utilizing only a beam splitter and photon number measurement.
However, the implementation of this test is still difficult due to the necessity of a photon number detection.
We can further simplify the measurement using threshold detectors for single-photon input states.
Since the output of the HBS is represented by the following states in the Fock basis: $\ket{2}\ket{0}$, $\ket{1}\ket{1}$, and $\ket{0}\ket{2}$, 
a single photon exists along a path only when the coincidence detection is observed by the threshold detectors placed at each output port of the HBS.
If the two input states are identical, no coincidences will occur, which implies that $P(1)=0$. 
This procedure is simply a HOM interference. 
Therefore, the SWAP test is equivalent to the HOM measurement for single photon states. We refer to this measurement, which characterizes indistinguishability, as the HOM test.

\subsection{HOM test for phase-randomized weak coherent pulses}
In the following, we will consider HOM interference of phase-randomized weak coherent pulses (WCP)\cite{chen_hongoumandel_2016,moschandreou_experimental_2018}, which are often utilized in QKD devices.
A coherent state of mean photon number $\mu$ and phase $\varphi$ can be written with the $k_p$-th basis of the mode $p$ as
\begin{equation}
    \ket{\phi} = \bigotimes_p \bigotimes_{k_p} \ket{\sqrt{\mu} e^{i \varphi}f_p (k_p)}_{k_p} ,
\end{equation}
where the function $f_p (k_p)$ stands for the relative amplitude of the $k_p$ state in the mode $p$. 
The function $f_p (k_p)$ satisfies
\begin{equation}
    \sum_{k_p} |f_p (k_p)|^2 = 1
\end{equation}
The HOM test will be performed by injecting two different pulses into the HBS\cite{HBT}.
We can describe one pulse state entering port a of the HBS as
\begin{equation}
    \ket{\phi}_\mathrm{a} = \left(\bigotimes_p \ket{\sqrt{\mu_a} }_{s_p}\right) \otimes \left(\bigotimes_p \bigotimes_{k_p \ne s_p} \ket{0}_{k_p}\right)
\end{equation}
by defining basis states appropriately. 
The pulse state entering the other port b may contain components other than $s_p$.
By collectively representing the Hilbert space associated with these components as $\perp$, the states of two different pulses can be expressed as follows without losing generality.
\begin{align}
    \ket{\phi}_\mathrm{a} &=\ket{\sqrt{\mu_\mathrm{a}}}_\mathrm{s} \otimes \ket{0}_\perp \nonumber\\
    \ket{\psi}_\mathrm{b}
    &= \ket{\sqrt{\mu_\mathrm{b}} e^{i \theta} \cos \Theta}_\mathrm{s} \otimes \ket{\sqrt{\mu_\mathrm{b}} e^{i \theta} \sin \Theta}_\perp
\end{align}
Here, we assume that the state $\ket{\phi}_\mathrm{b}$ has a random relative phase $\theta$ to $\ket{\phi}_\mathrm{a}$.  
The state $\ket{\phi}_\mathrm{b}$ contains a component $\ket{\sqrt{\mu_\mathrm{b}} e^{i \theta} \sin \Theta}_\perp$ in a Hilbert space orthogonal to that of $\ket{\sqrt{\mu_\mathrm{a}}}_\mathrm{s}$. 
The parameter $\Theta$ characterizes the overlap between the two states in the non-encoded degrees of freedom, with 
$\Theta =0$ corresponding to perfect indistinguishability. 
The output states of the HBS are also coherent states represented by 
\begin{align}
    \ket{\phi}_\mathrm{c} &=\ket{\alpha_{\mathrm{c}}}_\mathrm{s} \otimes \ket{\beta_{\mathrm{c}}}_\perp \nonumber\\
   \ket{\phi}_\mathrm{d} &=\ket{\alpha_{\mathrm{d}}}_\mathrm{s} \otimes \ket{\beta_{\mathrm{d}}}_\perp ,
\end{align}
where amplitudes $\alpha$ and $\beta$ are defined as follows:
\begin{align}
    \alpha_{\left\{\substack{\mathrm{c} \\ \mathrm{d}}\right\}} &= \frac{1}{\sqrt{2}}(\sqrt{\mu_\mathrm{a}} \pm \sqrt{\mu_\mathrm{b}} e^{i \theta} \cos \Theta) \nonumber\\
    \beta_{\left\{\substack{\mathrm{c} \\ \mathrm{d}}\right\}} &= \mp\frac{1}{\sqrt{2}}\sqrt{\mu_\mathrm{b}} e^{i \theta} \sin \Theta.
\end{align}
The probability $P_{mn}$ that $m$ and $n$ photons emerge at output ports c and d, respectively, is expressed by
\begin{align}
    P_{mn} &= \left(\sum_{k=0}^m e^{-|\alpha_\mathrm{c}|^2}\frac{|\alpha_\mathrm{c}|^{2k}}{k!}e^{-|\beta_\mathrm{c}|^2}\frac{|\beta_\mathrm{c}|^{2(m-k)}}{(m-k)!}\right)
    \left(\sum_{k=0}^n e^{-|\alpha_\mathrm{d}|^2}\frac{|\alpha_\mathrm{d}|^{2k}}{k!}e^{-|\beta_\mathrm{d}|^2}\frac{|\beta_\mathrm{d}|^{2(n-k)}}{(n-k)!}\right) \nonumber\\
    &=e^{-(\mu_\mathrm{c}+\mu_\mathrm{d})}\frac{\mu_\mathrm{c}^m \mu_\mathrm{d}^n}{m! n!},
\end{align}
where average photon numbers $\mu_{\mathrm{c}}$ and $\mu_{\mathrm{d}}$ are defined by
\begin{align}
    \mu_{\left\{\substack{\mathrm{c} \\ \mathrm{d}}\right\}}
    = (\mu_\mathrm{a} + \mu_\mathrm{b})/2 \pm \sqrt{\mu_\mathrm{a} \mu_\mathrm{b}} \cos \theta \cos \Theta.
\end{align}
Then, the coincidence probability in the HOM measurement is expressed by
\begin{align}\label{eq:coinc}
    P^{(\mathrm{coninc.})} &= 1 - \sum_{n=0}^\infty P_{0n} - \sum_{m=0}^\infty P_{m0} +P_{00} \nonumber\\
    &= 1 - e^{-\mu_\mathrm{c}} - e^{-\mu_\mathrm{d}} + e^{-(\mu_\mathrm{a}+\mu_\mathrm{b})},
\end{align}
where the last term reflects energy conservation in an HBS as $\mu_\mathrm{a}+\mu_\mathrm{b}=\mu_\mathrm{c}+\mu_\mathrm{d}$.
The expectation value of the coincidence probability is obtained by averaging Eq. (\ref{eq:coinc}) over the phase $\theta$.
\begin{align}
    \left\langle P^{(\mathrm{coninc.})} \right\rangle_\theta 
    &=1+ e^{-(\mu_\mathrm{a}+\mu_\mathrm{b})}-2 e^{-(\mu_\mathrm{a}+\mu_\mathrm{b})/2} I_0 (\sqrt{\mu_\mathrm{a} \mu_\mathrm{b}} \cos \Theta),
\end{align}
where $I_0(z)$ stands for the 0-th order modified Bessel function defined by
\begin{equation*}
    I_0(z)=\frac{1}{2 \pi} \int_0^{2 \pi} e^{\pm z \cos \theta} d\theta.
\end{equation*}
The visibility of the HOM interference is defined by
\begin{align}
    V_{\mathrm{HOM}} &= 1- \frac{\left\langle P^{(\mathrm{coninc.})} \right\rangle_\theta }{P^{\mathrm{c}} P^{\mathrm{d}}} \nonumber\\
    &=\frac{2 e^{-(\mu_\mathrm{a}+\mu_\mathrm{b})/2} (I_0 (\sqrt{\mu_\mathrm{a} \mu_\mathrm{b}}\cos \Theta)-1)}{(1-e^{-(\mu_\mathrm{a}+\mu_\mathrm{b})/2})^2},
\end{align}
where $P^{\mathrm{c}}=P^{\mathrm{d}}=1-\exp[-(\mu_\mathrm{a}+\mu_\mathrm{b})/2]$ represent the probabilities that photons appear at port c and port d, respectively, in the absence of two-photon interference. 

For a weak input ($\mu_{\mathrm{a}}, \mu_{\mathrm{b}} \ll 1$), the visibility is simplified by  approximating $I_0(z) \sim 1+z^2/4$ as
\begin{equation}\label{eq:VHOM_app}
     V_{\mathrm{HOM}} \approx \frac{2\mu_\mathrm{a}\mu_\mathrm{b}\cos^2 \Theta}{(\mu_\mathrm{a}+\mu_\mathrm{b})^2} \le \frac{1}{2}
\end{equation}
The equality holds when $\mu_\mathrm{a}=\mu_\mathrm{b}$ and $\cos^2 \Theta =1$.
If the state associated with $\ket{\beta}_\perp $ is mixed, $x= \cos^2 \Theta$ in Eq. (\ref{eq:VHOM_app}) is a probabilistic variable.  
The mean and variance of the HOM visibility are then given by the mean and variance of $x$, respectively.

The HOM visibility can be related to the expected results of the SWAP test as follows.
Equation (\ref{eq:Kerr}) implies that no phase shift is applied when even number of photons emerge at the output port c, and the outcome of the SWAP test is "0". Conversely, the phase shift of $\pi$ is applied when an odd number of photons is present, leading to an outcome of "1".
Therefore, the probabilities of obtaining the outcomes "0" and "1" are given by 
\begin{align*}
    P(0) &= \sum_{k=0}^\infty \sum_{n=0}^\infty P_{2k,n} 
    = \frac{1+e^{-2\mu_\mathrm{c}}}{2} \nonumber\\
     P(1) &= \sum_{k=0}^\infty \sum_{n=0}^\infty P_{2k+1,n} 
     = \frac{1-e^{-2\mu_\mathrm{c}}}{2},
\end{align*}
which results in
\begin{equation}
    \sqrt{\frac{P(0)-P(1)}{P(0)+P(1)} }=e^{-(\mu_\mathrm{a} + \mu_\mathrm{b})/2} e^{\sqrt{\mu_\mathrm{a} \mu_\mathrm{b}} \cos \theta \cos \Theta}.
\end{equation}
Since fidelity between the states $\ket{\phi}_\mathrm{a}$ and $\ket{\psi}_\mathrm{b}$ is given by
\begin{align}
    F\left(\ket{\phi}_\mathrm{a}, \ket{\psi}_\mathrm{b}\right) &= \left|\braket{\sqrt{\mu_\mathrm{a}}|\sqrt{\mu_\mathrm{b}}e^{i\theta} \cos \Theta}_\mathrm{s}\braket{0|\sqrt{\mu_\mathrm{b}}e^{i\theta} \sin \Theta}_\perp \right| \nonumber\\
    &= e^{-(\mu_\mathrm{a} + \mu_\mathrm{b})/2} e^{\sqrt{\mu_\mathrm{a} \mu_\mathrm{b}} \cos \theta \cos \Theta},
\end{align}
we again obtain Eq. (\ref{eq:fidelity}) for WCP. 
However, as $\mu_\mathrm{a}$ and $\mu_\mathrm{b}$ approach zero, the fidelity tends to unity because any coherent states become the vacuum state. 
To evaluate the distinguishability due to the effects of other degrees of freedom, we should exclude the contribution of the vacuum, which corresponds to $\exp[-(\mu_\mathrm{a} + \mu_\mathrm{b})/2]$. 
 We define the non-zero photon contribution to the fidelity as
 \begin{align}\label{eq:F_tilde}
     \tilde{F} \left(\ket{\phi}_\mathrm{a}, \ket{\psi}_\mathrm{b}\right) 
     &= \frac{1}{\sqrt{P^\mathrm{c} P^\mathrm{d}}}\left| e^{-(\mu_\mathrm{a} + \mu_\mathrm{b})/2} \left(\exp[\sqrt{\mu_\mathrm{a} \mu_\mathrm{b}} e^{i \theta} \cos \Theta] -1 \right) \right|.
 \end{align}
 Squaring Eq. (\ref{eq:F_tilde}) and averaging over the random phase $\theta$ yields 
\begin{align}\label{eq:fidelity-VHOM}
    \left\langle  \left|\tilde{F} \left(\ket{\phi}_\mathrm{a}, \ket{\psi}_\mathrm{b}\right) \right|^2\right\rangle_\theta &= \frac {e^{-(\mu_\mathrm{a} + \mu_\mathrm{b})}}{\left(1 - e^{-(\mu_\mathrm{a} + \mu_\mathrm{b})/2} \right)^2} \left(I_0( 2 \sqrt{\mu_\mathrm{a} \mu_\mathrm{b}} \cos \Theta ) -1 \right)\nonumber\\
    &\approx  \frac{4\mu_\mathrm{a}\mu_\mathrm{b}\cos^2 \Theta}{(\mu_\mathrm{a}+\mu_\mathrm{b})^2} = 2 V_\mathrm{HOM}.
\end{align}
We have established a connection between fidelity and HOM visibility, which allows us to perform HOM test as SWAP test for WCP. 
The HOM test is more effective than fidelity measurement in evaluating the effects of other degrees of freedom on indistinguishability, as it eliminates the influence of the vacuum contribution.
Generally, the states at ports a and b are mixed, so Eq. (\ref{eq:fidelity-VHOM}) should also be averaged over $\Theta$ with an appropriate probability distribution.  Then, LHS of Eq. (\ref{eq:fidelity-VHOM}) should be represented by $\mathrm{Tr}(\rho_a \rho_b)$.

\section{Experiment}\label{sec:experiment}
In this study, we performed the HOM test on signals from a transmitter that implements the BB84 protocol using time-bin encoding. 
We observed the HOM interference between the adjacent pulses, one of which was delayed by one clock cycle to investigate the indistinguishability of the pulses.
We encoded quantum signals in the X and Y bases with WCP as
\begin{align}
    \ket{X0} &= \left| \sqrt{\mu/2} \;e^{i \varphi} \right\rangle_{t_0}\left| \sqrt{\mu/2} \;e^{i \varphi}\right\rangle_{t_1} \nonumber\\
    \ket{X1} &= \left| \sqrt{\mu/2} \;e^{i \varphi} \right\rangle_{t_0}\left| -\sqrt{\mu/2} \; e^{i \varphi} \right\rangle_{t_1}  \nonumber\\
    \ket{Y0} &= \left| \sqrt{\mu/2} \; e^{i \varphi}\right\rangle_{t_0}\left|i \sqrt{\mu/2} \; e^{i \varphi} \right\rangle_{t_1} \nonumber\\
    \ket{Y1} &= \left| \sqrt{\mu/2} \; e^{i \varphi} \right\rangle_{t_0}\left|-i \sqrt{\mu/2} \; e^{i \varphi} \right\rangle_{t_1},     
\end{align}
where $\ket{\nu}_t$ stands for a WCP state centered at time $t$ with an average photon number $\nu$. 
Phase $\varphi$ varies randomly with each pulse. 
The time-bin components are separated enough to be orthogonal as ${}_{t_0}\braket{\nu|\nu}_{t_1}=0$.
We measured the HOM interference signal at time either $t_0$ or $t_1$, which corresponds to the Z-basis measurement.
This Z-basis measurement eliminates the dependence of the interference on the quantum signal states encoded in X-basis and Y-basis.

\begin{figure}[htbp]
\centering\includegraphics[width=\linewidth]{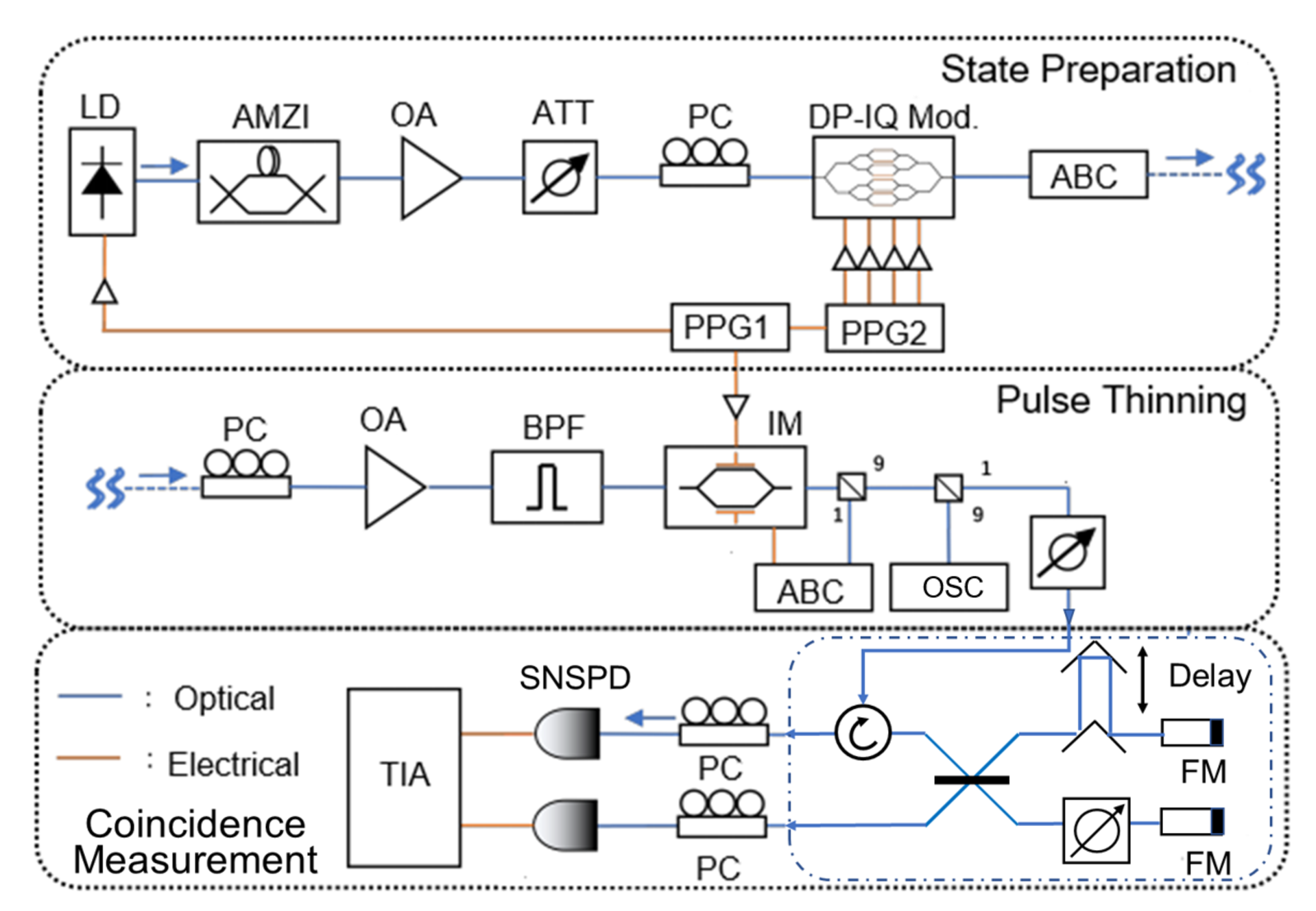}
\caption{Experimental setup for HOM test. The top, middle, and bottom represent the state preparation part, the pulse thinning part, and the coincidence measurement part, respectively. LD: Laser Diode, AMZI: Asymmetric Mach-Zehnder Interferometer, OA: Optical Amplifier, ATT: Attenuator, PC: Polarization Controller, ABC: Automatic Bias Controller, PPG: Pulse Pattern Generator, BPF: Band Pass Filter, OSC: Oscilloscope, FM: Faraday Mirror, SNSPD: Superconducting Nanowire Single Photon Detector, TIA: Time Interval Analyzer. }
\label{fig:setup}
\end{figure}

Figure \ref{fig:setup} shows the experimental setup. 
The whole setup was constructed using off-the-shelf fiber-optic components.
The top shows a transmitter that prepares BB84 states. A laser diode (LD) operating in gain-switched mode emitted optical pulses with a duration less than 50 ps and a repetition rate of 1.25 GHz.
An asymmetric Mach-Zehnder interferometer (AMZI) converted the optical pulses to double-pulses separated by 400 ps. 
The double-pulses were modulated into one of the BB84 states in X- or Y-basis.
The modulation scheme was an expansion of the one described in our previous report\cite{zhang_state_2020}, where an In-phase and Quadrature (IQ) modulator (or a dual-parallel modulator) was used to provide precise qubit states against the distortion of electrical drive signals in high-speed QKD transmitters. 
The present implementation employed a commercially available dual-polarization IQ modulator, where the incident light is split into two by polarization and modulated by two IQ modulators. By combining the polarized components appropriately, the intensity can be modulated simultaneously with high precision for the decoy method. In this experiment, we fixed the intensity of the pulses. 
We employ an automatic bias controller (ABC) to compensate for the drift of the modulator.

The middle illustrates the pulse-thinning part. 
Single-photon detectors are blind for a recovery time after photon detection, which hinders the accurate detection of coincidence events.
To obtain precise coincidence counts, we thinned the pulse train. An intensity modulator (IM) selected a pulse at an interval of 51.2 ns, which was chosen to ensure that the recovery time (10 ns) did not affect the measurement outcomes.
The thinned-out pulses were then monitored using an oscilloscope.
We set the average photon number of the pulses to 0.5 photons/pulse at the output of the pulse-thinning part.
Er-doped fiber amplifiers (labeled OA in Fig. \ref{fig:setup}) were used to compensate for loss in the modulators. 
Amplified spontaneous emission noise from the amplifiers was filtered out using a 100-GHz bandwidth bandpass filter (BPF).

The bottom illustrates the coincidence measurement to observe HOM interference between the adjacent pulses. 
An asymmetric Michelson interferometer provided a path difference $T'=T+\tau$, where $T=800$ ps corresponds to one clock cycle.
In the present experiment, $\tau$ was nominally selected from the set $\tau \in \{-26$  ps, -16 ps, -6 ps, -4 ps, -2 ps, 0 ps, 2 ps, 4 ps, 6 ps, 16 ps, 26 ps\}.
In the interferometer, Faraday mirrors were used to compensate for the polarization disturbance in the optical paths. An attenuator was inserted in one path to equalize the loss in the two paths.
Two Superconducting Nanowire Single Photon Detectors (SNSPDs) detect output photons from the interferometer. Since the detection efficiencies of SNSPDs exhibit strong polarization dependence, polarization controllers (PCs)  were placed at the inputs of SNSPDs to optimize the polarization of input photons. 
Finally, a time interval analyzer (TIA) measured the coincidence of the photon detection signals from the two detectors. 
The time window for photon detections was selected to facilitate the count of photons belonging to one of the double-pulse components, which corresponded to the Z-basis measurement.
The coincidence window of the TIA was set to 200 ps, taking into account the specifications of the time analyzing apparatus, that is, the time resolution and jitter of the TIA (100 ps and 28 ps, respectively), and the jitters of two SNSPDs (24 ps and 14 ps).

When measuring the HOM interference between adjacent pulses, one pulse state was fixed at X0 while the adjacent pulse state varied between X1, Y0, and Y1.
Additionally, HOM interference was measured between unmodulated pulses for reference, which was expected to yield maximum visibility.
The coincidence counts were measured for 30 seconds at each path-difference value $\tau$. 
The measurements were repeated five times, and the average ($\{y_i\}$) and standard deviation ($\{s_i\}$) of the coincidence counts were calculated.

\section{Results}\label{sec:results}
\begin{figure}[htbp]
    \centering\includegraphics[width=0.8\linewidth]{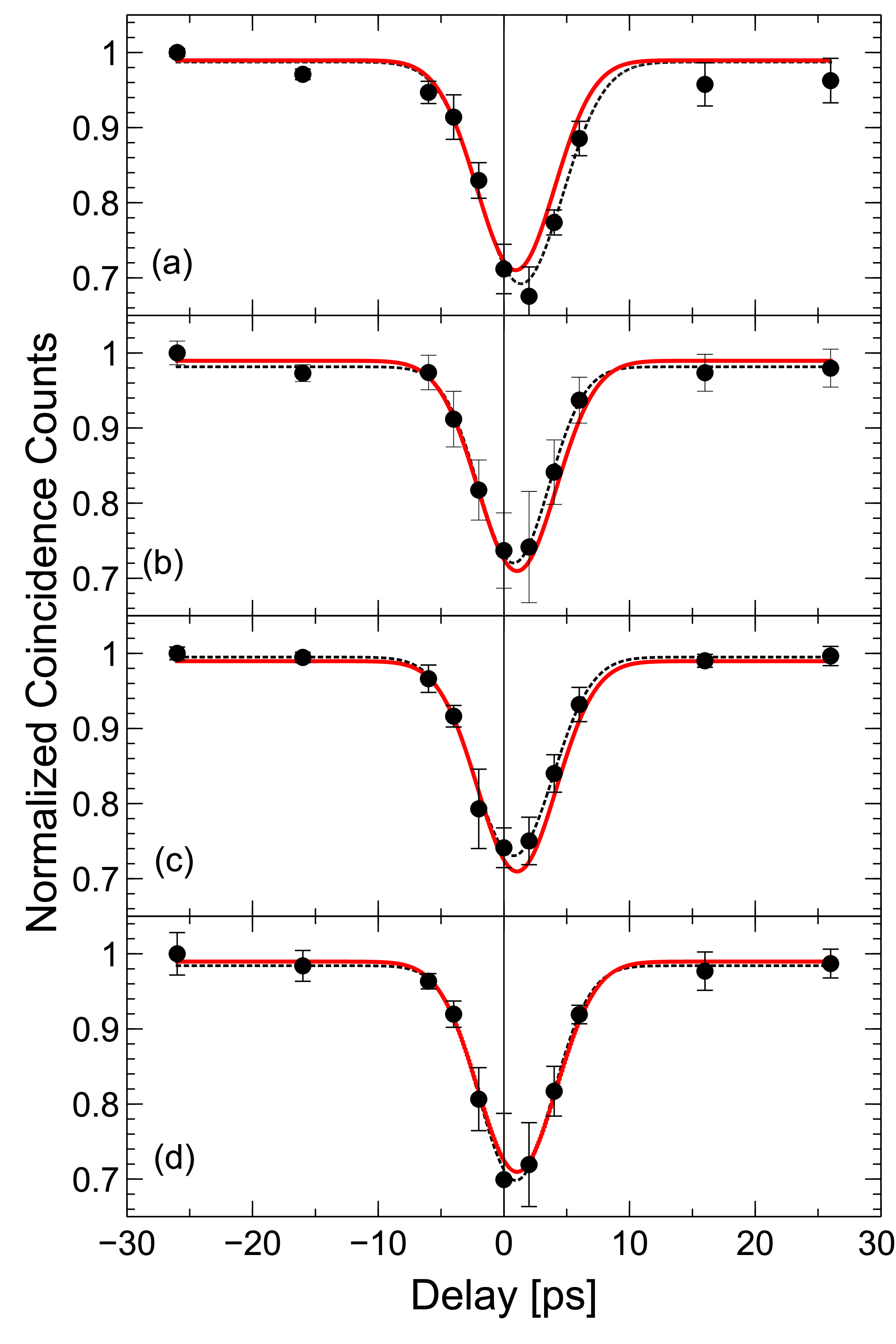}
    \caption{Results of Hong-Ou-Mandel interference between adjacent pulses: (a) unmodulated pulses for reference, (b) pulses in X1-state and X0-state, (c) Y0-state and X0-state, and (d) Y1-state and X0-state. Coincidence counts were normalized using the value obtained at -26 ps. Red solid lines represent fitting results with a single set of parameters, while dotted lines represent fitting results using optimized parameters for each measurement group. }
    \label{fig:HOM_results}
\end{figure}

Figure \ref{fig:HOM_results} shows the results of the HOM interference experiment between pulses.
The coincidence counts were then normalized so that the value was set to unity at -26 ps, well separated from the dip. 
The coincidence count shows a clear dip near 0 ps, indicating HOM interference.
The following observations were drawn from the measurement results:
The visibility was about 0.3, which is smaller than the theoretical maximum value of 0.5. 
However, no significant differences were observed between the measurement groups.  
The measured coincidence fluctuated more significantly near the dip than in the tails. 
The dip positions also show only a slight variation.
The dip widths were 7–8 ps (FWHM),  smaller than the pulse width of 30–50 ps.

To conduct a detailed analysis, we fitted the data to the function as
\begin{align}\label{eq:fitting}
    y_{i,j} &= 1 - f(t_i; A_j, t_{0,j}, \sigma_j ,B_j) \\
    f(t; A, t_0, \sigma ,B) & = A \exp\left[-\frac{(t-t_0)^2}{2\sigma^2}\right] + B,
\end{align}
using the curve\_fit function from the SciPy library\cite{SciPy-NMeth2020}.
The fitting was weighted by the standard deviations of the coincidence counts.
The dataset $\{y_{i,j}\}$ represents the normalized coincidence counts at time $t_i$ for the measurement group $j \in \{\mathrm{unmodulated, X1\text{-}X0, Y0\text{-}X0, Y1\text{-}X0}\}$. 
The visibility of the HOM interference was estimated by
\begin{equation}\label{eq:vis_exp}
    V_j = 1- \frac{I_{\mathrm{min},j}}{I_{\mathrm{max},j}}=1-\frac{1-(A_j+B_j)}{1-B_j},
\end{equation}
where the standard deviation was given by
\begin{equation}
    \sigma_{V,j}^2 = \sigma_{A,j}^2 + \sigma_{B,j}^2.
\end{equation}
Table\ref{tab:fitting} shows the optimized fitting parameters with uncertainties (standard deviations). 
The results of fitting were also shown in Fig. \ref{fig:HOM_results}

\begin{table}[htbp]
    \caption{Values and standard deviations of the optimized fitting parameters for the observed HOM interference. Parameters in "whole" were obtained by fitting all the observed data with a single set of the parameters, while those in "unmodulated", "X1 - X0", "Y0-X0", and "Y1-X0" were obtained by fitting independently for each group. }
    \label{tab:fitting}
    \centering
   \begin{tabular}{lcccccc}
        \hline
         & V     & std ($V$) & $t_0$ [ps]    & std ($t_0$) & $\sigma$ [ps] & std ($\sigma$)   \\ \hline
        unmodulated & 0.299 & 0.023 & 1.36 & 0.21  & 3.28 & 0.27      \\ \hline
        X1-X0    & 0.266 & 0.042 & 0.773 & 0.40  & 2.83 & 0.39       \\ \hline
        Y0-X0    & 0.266 & 0.022 & 0.767 & 0.25  & 3.11 & 0.23       \\ \hline
        Y1-X0    & 0.291 & 0.042 & 0.943 & 0.27  & 2.94 & 0.28     \\ \hline
        whole    & 0.283 & 0.014 & 1.05 & 0.12 & 3.12 & 0.12      \\ \hline
    \end{tabular}
\end{table}

We first statistically investigated the HOM visibilities, as they are a crucial metric for assessing the indistinguishability of pulses. 
The estimated values of the visibility ranged from 0.266 to 0.299.
 The differences between any two visibilities, $V_i$ and $V_j$, were smaller than the standard deviation estimated by $\mathrm{std}_{ij}^2=\mathrm{std}_i^2+\mathrm{std}_j^2$ for Gaussian distributions.  
We compared the likelihood of the multi-parameter set and that of a single parameter set defined by
\begin{align*}
    L_{\mathrm{multi}} &= \prod_{j} \prod_{i=1}^N \frac{1}{\sqrt{2 \pi} s_i}\exp \left[-\frac{(y_i-f(t_i,\mathbf{\theta}_j))^2}{2 s_i^2}\right] \\
     L_{\mathrm{single}} &= \prod_{j} \prod_{i=1}^N \frac{1}{\sqrt{2 \pi} s_i}\exp \left[-\frac{(y_i-f(t_i,\mathbf{\theta}_{\mathrm{s}}))^2}{2 s_i^2}\right], 
\end{align*}
where 
the parameter set $\mathbf{\theta}_j =\{A_j, t_{0,j}, \sigma_j ,B_j\}$  was obtained independently for the group $j$ and the set $\mathbf{\theta}_{\mathrm{s}} =\{A_{\mathrm{s}}, t_{0,\mathrm{s}}, \sigma_{\mathrm{s}}, B_{\mathrm{s}}\}$ was obtained by applying a single parameter set for whole data. 
The log-likelihood ratio was calculated  as
\begin{equation}
    LR = \sum_j \sum_{i=1}^N \left(\frac{(y_i-f(t_i,\mathbf{\theta}_j))^2}{s_i^2}-\frac{(y_i-f(t_i,\mathbf{\theta}_{\mathrm{s}}))^2}{s_i^2}\right)
\end{equation}
for the likelihood ratio test\cite{Akaike1998,casella_statistical_2024}. 
In this test, the degree of freedom is $12 = 4 \; \mathrm{parameters} \times 4 \; \mathrm{groups} - 4 \; \mathrm{parameters} \times 1 \; \mathrm{groups}$.
The test resulted in a $p$-value of $0.18$, which holds the null hypothesis that the coincidence counts were sampled from the same population at the 5 \% significance level. 
We also estimated the statistical power of our test simply assuming a Gaussian distribution.  
It was found that a visibility difference of $\Delta V > 0.05$ would be detected 
with 80 \% probability given our measurement precision. 
The observed differences ($|\Delta V| < 0.035$) are below this 
detection threshold, which supports the hypothesis on indistinguishability of modulated photon pulses.

It is necessary to consider the effects of deterministic time shifts by encoding quantum states.  
If modulation affects pulse timing, a fixed time shift only moves the dip position and does not affect visibility. In contrast, fluctuations in the time shift will diminish visibility.
The observed positions of the dips varied by less than 0.6 ps, which is smaller than the width of the dips (7-8 ps), regardless of the modulation.
We performed a one-way analysis of variance (ANOVA)\cite{Gillard2020_ANOVA} to determine if there were any statistically significant differences in the means of the following groups: unmodulated, X1-X0, Y0-X0, and Y1-X0.  
The ANOVA yielded a $p$-value of 0.2, indicating that we cannot reject the null hypothesis that the mean positions of the dips are the same at the 5 \% significance level. 
In the present analysis, we assumed that the dip position deviated at the transmitter. 
In practice, however, the time origin of the asymmetric Michelson interferometer in the coincidence measurement (the bottom part in Fig. \ref{fig:setup}) likely shifted for the measurement of each state. 
Considering this and adjusting the time origin, the modulation-induced difference in HOM interference will be even smaller.

\section{Discussion}\label{sec:discussion}
The experimental results on HOM interference indicate that modulation had a negligible effect on the visibility. 
Therefore, the indistinguishability of the pulses is independent of the encoding in the present system. 
However, it should be noted that there is still a possibility of a type II error, meaning that we may fail to reject the false null hypothesis considered in Sec. \ref{sec:results}.
To improve statistical power, it is necessary to reduce the coincidence count fluctuation and decrease the variance of the fitting parameters. 
Extending the measurement time could reduce statistical errors arising from finite data.
Then, the observed fluctuation is determined by the variance of $x= \cos^2 \Theta$ in Eq. (\ref{eq:VHOM_app}), according to the degree of mixing of the light states. 
Even if light is in a mixed state, it does not affect security as long as the state is independent of modulation. In other words, as long as the state originates from the light source, it does not affect security.

Since a mixed state reduces the HOM visibility, it has been studied in the context of Measurement-Device-Independent (MDI) QKD\cite{lo_measurement-device-independent_2012,ferreira_da_silva_proof--principle_2013,ferreira_da_silva_long-distance_2013,rubenok_real-world_2013,liu_experimental_2013,tang_experimental_2014,yuan_interference_2014,ge_analysis_2023}. 
Unlike MDI-QKD, a decrease in the visibility of the HOM interference does not affect the validity of the HOM test for indistinguishability. This is because the test verifies that modulation does not affect the interference. However, decreased visibility makes changes to the HOM interference more difficult to observe, so it is preferable for the interference to be clearly visible.

In experiments, a gain-switched laser diode may emit light with varying spectral property, intensity, and temporal behavior for each pulse, resulting in a mixed state of the light.
Yuan \textit{et al.}\cite{yuan_interference_2014} suggested that the resulting spectral mismatch significantly affects HOM visibility. 
By limiting the spectral filter bandwidth from 2 THz to 72.5 and 13.8 GHz, they observed an increase in visibility from 0.25 to 0.28 and 0.46, respectively.
Because the present experiment employed a 100-GHz bandwidth spectral filter, the measured HOM visibility values are consistent with their observations. 
Jitter will also reduce visibility.
We assume two Gaussian pulses with variance $\sigma^2$ and peak difference $d$. We further assume that $d$  follows a normal distribution with mean 0 and variance $s^2$. 
The overlap between the pulses is known to be erfc$(|d|/(2 \sqrt{2} \sigma))$.
The expectation value can be described by 
\begin{equation}
    I_{\mathrm{jitter}} = \frac{2}{\pi} \arctan \left(\frac{2 \sigma}{s} \right).
\end{equation}
The reported jitter values for gain-switched DFB laser diodes are 2.2 ps\cite{schell_low_1994} and 5-6 ps\cite{weber_measurement_1992}.
The estimated effect of jitter is $I_{\mathrm{jitter}} =0.85-0.96$ for a pulse duration of $\tau_\mathrm{p} = 2 \sqrt{2 \ln 2} \sigma = 30-50$ ps in this experiment. 
Therefore, the above characteristics of the gain-switched DFB laser diode can explain the decrease in the observed HOM visibility from its ideal value.
This observation is consistent with the experimental results that interference was independent of modulation.

The effect of intensity fluctuation is small.
If we assume that a small fluctuation in intensities of the two pulses as $\mu_i = \mu_0 (1+\delta_i)$; $i \in \{\mathrm{a},\mathrm{b}\}$, where $\delta_i$ obeys a normal distribution with a mean $0$ and a variance $s_I^2$, the factor $\mu_\mathrm{a} \mu_\mathrm{b}/(\mu_\mathrm{a} + \mu_\mathrm{b})^2$ in Eq. (\ref{eq:VHOM_app}) is reduced to 
\begin{equation}
    \left\langle \frac{ \mu_\mathrm{a} \mu_\mathrm{b}}{(\mu_\mathrm{a} + \mu_\mathrm{b})^2} \right\rangle = \frac{1}{2}\left(1-\frac{1}{2} s_I^2 \right) + O(s_I^4)
\end{equation}
with a variation
\begin{equation}
    V^2\left( \frac{ \mu_\mathrm{a} \mu_\mathrm{b}}{(\mu_\mathrm{a} + \mu_\mathrm{b})^2} \right) = \frac{1}{4} s_I^4 + O(s_I^6).
\end{equation}
The fluctuation of a gain-switched laser diode was observed to be $s_I^2 = 7 \times 10^{-4}$, for a high drive current\cite{Nakata_intensity_2017}. 
Therefore, visibility decreases by 0.4 \%. Increase in fluctuation will be negligible.


The imperfections in the measurement system also result in reduced visibility and increased fluctuations.
The imperfections include saturation in the detectors, imbalance in detection efficiencies and dark counts, and imbalance in losses in the interferometer \cite{thomas_measurement_2010,wang_realistic_2017,kim_two-photon_2021}. 
However, as shown below, the imperfections have a limited impact on the present experiment.
In this experiment, we employed SNSPDs with sufficiently sparse pulse inputs. 
Therefore, saturation effects in the detectors are negligible. 
Efficiencies and dark counts of the SNSPDs were around 63 \% and 20 Hz, respectively. 
Thus, the dark counts had a negligible impact on the measurement outcomes.
The effect of the imbalance of the interferometer should be small because an attenuator was placed to compensate for the imbalance in the experiment. 
Nevertheless, we cannot entirely eliminate the effects of drift and fluctuation in the experimental devices. Further improvement is desirable.


The spectral-temporal characteristics of the gain-switched laser diodes also affect the width of the HOM dip\cite{legero_time-resolved_2003,kim_time-resolved_2021}.
If the pulses were not transform-limited, the observed width of the HOM dip is smaller than the pulse duration.
Suppose that the complex amplitude of a chirped pulse is written by $A(t)=A_0 \exp[-(1-ia)t^2/\tau_a^2]$, where the chirp is characterized by a chirp parameter $a$\cite{saleh_fundamentals_2013,chen_coherence_2013}. 
A real time constant $\tau_a$ is related to the pulse duration $\tau_\mathrm{p}$ (FWHM) by $\tau_\mathrm{p}=\sqrt{2 \ln 2}\; \tau_a$.
The spectral intensity $\mathcal{S}(\nu)$ is expressed by
\begin{equation}
    \mathcal{S}(\nu) \propto \exp \left[-\frac{2 \pi^2 \tau_a^2 (\nu-\nu_0)^2}{1+a^2}\right],
\end{equation}
which is a Gaussian function of frequency $\nu$ with FWHM $\Delta \nu = (\sqrt{2 \ln 2}/\pi)(\sqrt{1+a^2}/\tau_a)=(2 \ln 2 /\pi)(\sqrt{1+a^2}/\tau_\mathrm{p})$.
Since the HOM interference is described by 
\begin{align*}
    S_{\mathrm{HOM}}(t) &=1-\frac{1}{2}\exp\left[-\frac{\pi^2}{\ln 2} \Delta \nu^2 (t-t_0)^2 \right]\\
    &= 1-\frac{1}{2}\exp\left[-\frac{4 \ln 2\;(1+a^2)}{\tau_\mathrm{p}^2} (t-t_0)^2 \right],
\end{align*}
the chirp parameter can be obtained from the pulse duration and the width of the HOM dip by comparing it with Eq.(\ref{eq:vis_exp}) as
\begin{equation}
    a= \sqrt{\frac{\tau_\mathrm{p}^2}{(8 \ln 2) \sigma^2}-1},
\end{equation}
where a small parameter $B$ is neglected for simplicity.
Using experimentally obtained values $\sigma^2 \approx 10$ ps and $\tau_{\mathrm{p}} =30-50$ ps, the estimated value of the chirp parameter is between 4 and 7, which is consistent with those reported in the literature\cite{agrawal_nonlinear_1995,consoli_time_2011}.
It should be noted that the above analysis is based on the linearly chirped Gaussian spectrum.
In order to extract quantitative information on chirping from the HOM interference, we need to consider the detailed dynamics of gain-switched lasers.

\section{Conclusion}\label{sec:conclusion}
The indistinguishability of transmitted pulses is a critical assumption in the security of the QKD protocol. 
We have developed an indistinguishability test method based on the HOM dip that appears in two-photon interference between pulses. 
This HOM test offers an effective method for examining indistinguishability without relying on the qubit assumption by considering all degrees of freedom in a simple experimental setup.
The feasibility of the HOM test was demonstrated by observing a distinct interference dip using pulses from a QKD transmitter that implemented the decoy-BB84 protocol.
The shapes and visibilities of the observed HOM dips were found to be independent of the BB84 state selection within experimental error.
This finding indicates that the indistinguishability assumption is satisfied in the transmitter setup under investigation.
We then analyzed the HOM dips, accounting for the mixed-state nature of the pulses. 
This analysis showed that the characteristics of these pulses could explain the experimentally observed depths and widths of HOM dips.
A thorough future analysis will provide valuable insights into the pulse states.
The present method can be applied to other encoding schemes, such as polarization, by placing a polarization analyzer in front of the photon detectors. 

Since HOM visibility is unaffected by fidelity enhancement due to vacuum contributions for weak coherent light, this method more clearly reflects state differences than the fidelity method. 
However, for mixed states, HOM visibility measures the trace of the product of the density matrices, rather than the fidelity directly. 
HOM interference becomes weaker in mixed states, leading to reduced measurement precision. 
For high-precision discrimination, higher visibility is preferable, so a light source closer to a pure state is desirable. 
This experiment achieved a visibility of about 0.3, which is considered sufficient for the indistinguishability test. 
On the other hand, since the HOM test can measure the fidelity between a pure state and a mixed state, it is well-suited for measuring the distance of a mixed state from a pure reference state.

In conclusion, this approach can be used to facilitate a practical evaluation of security certification for QKD systems and will also promote the standardization of QKD system evaluation.

\begin{appendices}
\setcounter{equation}{0}
\renewcommand{\theequation}{A-\arabic{equation}}
\section{Indistinguishability expressed by the trace of the density operator product}\label{sec:Appendix}

The result of the SWP test is given by the trace of the density operator product: 
 $\mathrm{Tr}(\tilde{\rho_1} \tilde{\rho_2} )$,
where the density operators $\tilde{\rho}_1$ and $\tilde{\rho}_2$ represent the states of adjacent pulses. The states are represented by the tensor product of the state of the degree of freedom (DOF) that carries bit and basis information and the state of other DOFs:
\begin{equation}
    \tilde{\rho}_i = \rho_{S_i} \otimes \rho_i, \quad i=1,2   
\end{equation}
Suppose BB84 protocol is implemented with X and Y bases. Then, the HOM visibility reflects the trace of $\rho_1 \rho_2$ by applying the projection $\ket{\mathrm{Z0}}\bra{\mathrm{Z0}}$ or $\ket{\mathrm{Z1}}\bra{\mathrm{Z1}}$, because the projection results in the same states $\ket{\mathrm{Z0}}$ or $\ket{\mathrm{Z1}}$ irresptive to the prepared states $\ket{\mathrm{X0}}$ and $\ket{\mathrm{X1}}$ for X basis and $\ket{\mathrm{Y0}}$ and $\ket{\mathrm{Y1}}$ for Y basis.
Note that imperfect state preparation may alter the probability $\braket{\mathrm{Zi}|\rho_{S_j}|\mathrm{Zi}}$, which reduces the visibility.
This imperfection is mainly due to the imbalance of the AMZI. Therefore, the difference between the bases will be small.

The state of other DOFs can be written by the tensor product of the state in the Hilbert space of each DOF. We here introduce pure states $\ket{\lambda_k}$ with a single index $k$ running over the whole Hilbert space. We assume that $\{\ket{\lambda_k}\}$ is an orthonormal basis. Then, the pulse states of other DOFs are represented by the statistical mixtures of the basis states:
\begin{align}
    \rho_i = \sum_k p_{i,k} \ket{\lambda_k} \bra{\lambda_k}, 
\end{align}
where
\begin{align*}
  \sum_k p_{i,k} = 1 \\
  0 \le p_{i,k} \le 1.
\end{align*}
Note that even before the state preparation, the trace of the product may be smaller than unity, because
\begin{equation}
    \mathrm{Tr}(\rho_1 \rho_2) = \sum_k p_{1,k}p_{2,k} \le \sum_k p_{i,k}^2 \le 1
\end{equation}
The above inequality can be proven with Cauchy-Schwarz's inequality. The first equality holds if $p_{1,k}=p_{2,k}$ for all $k$, and the second holds if the states are pure. In other words, the trace of the product is less than unity, unless the adjacent pulses are in the same pure state. 

The effect of the modulation can be written with a unitary transformation $U_{S_i}$. The trace of the product of the density operators is given by
\begin{align}
    \mathrm{Tr}(U_{S_1}\rho_1 U_{S_1}^\dagger U_{S_2}\rho_2 U_{S_2}^\dagger) &= \mathrm{Tr}(U_{S_2}^\dagger U_{S_1}\rho_1 U_{S_1}^\dagger U_{S_2}\rho_2 ) \nonumber \\
    &= \mathrm{Tr}(U_{S}\rho_1 U_{S}^\dagger \rho_2) \nonumber\\
    & \le  \mathrm{Tr}(\rho_1 \rho_2), \label{eq_5}
\end{align}
where $U_{S}$ is defined by $U_{S}=U_{S_2}^\dagger U_{S_1}$.
The equality holds if the $U_{S}$ is the identity operator, i.e.,
\begin{equation}
    U_{S_1} = U_{S_2}
\end{equation}
The inequality (\ref{eq_5}) is proven as follows: a unitary operator transforms an orthonormal basis $\{\ket{\lambda_k}\}$ into another orthonormal basis $\{\ket{\mu_k}\}$. Then, the density operator $\rho_1$ is transformed as 
\[U_{S}\rho_1 U_{S}^\dagger = \sum_k q_{1,k} \ket{\mu_k}\bra{\mu_k}.\]
Therefore, 
\begin{equation}\label{eq_6}
   \mathrm{Tr}(U_{S}\rho_1 U_{S}^\dagger \rho_2) = \sum_k q_{1,k} p_{2,k} |\braket{\mu_k|\lambda_k}|^2 \le \sum_k q_{1,k} p_{2,k}, 
\end{equation}
because $|\braket{\mu_k|\lambda_k}|^2 \le 1$. The equality holds $\ket{\mu_k} = \ket{\lambda_k}$ for all $k$, that is, $U_{S}=\hat{1}$. Then, $q_{1,k} = p_{1,k}$ and RHS of (\ref{eq_6}) reads $\mathrm{Tr}(\rho_1 \rho_2) $

We have shown that the trace of the product of the modulated density operators 
\[\mathrm{Tr}(U_{S_1}\rho_1 U_{S_1}^\dagger U_{S_2}\rho_2 U_{S_2}^\dagger)
\]
is less than that of the unmodulated density operators $\mathrm{Tr}(\rho_1 \rho_2)$ unless the effects are identical for all the modulation patterns. Therefore, we can say that the indistinguishability is not compromised by the modulation if the visibility measurements on the modulated states show no difference from those on the unmodulated states.

\end{appendices}

\begin{backmatter}
\bmsection{Funding}
This study was partly supported by MIC under the initiative grant “Research and Development for
Construction of a Global Quantum Cryptography Network” (JPMI00316) and KAKENHI grant number 23K25793.

\bmsection{Acknowledgments}
The authors thank Prof. Kiyoshi Tamaki of Toyama University for fruitful discussions.

\bmsection{Disclosures}

\noindent The authors declare no conflicts of interest.

\bmsection{Data availability} Data underlying the results presented in this paper are not publicly available at this time but may be obtained from the authors upon reasonable request.

\end{backmatter}

\bibliography{quantum_crypto_references}

\end{document}